\documentclass[useAMS,usenatbib]{mn2e}
\usepackage{aurelie}
\usepackage{graphicx}
\usepackage{natbib}
\usepackage{pjournal}
\usepackage{rotating}
\usepackage{multirow}
\usepackage{IEEEtrantools}

\title[The HI bias] {A scale dependent bias on linear scales: the case for HI intensity mapping at $z = 1$}
\author[A. P\'enin et al.]{Aur\'elie P\'enin$^{1}$\thanks{E-mail: aurelie.c.penin@gmail.com (AVR)}, 
Obinna Umeh$^{2}$, 
Mario G. Santos$^{2,3}$ \\
$^{1}$Astrophysics and Cosmology Research Unit, School of Mathematics Statistics and Computer Science, \\
University of KwaZulu-Natal, Durban, 4041, South Africa\\
$^{2}$Department of Physics and Astrophysics, University of the Western Cape, Cape Town 7535, South Africa\\
$^{3}$SKA South Africa, The Park, Cape Town 7405, South Africa}

\begin{document}
\date{Accepted 0000. Received 0000}

\pagerange{\pageref{firstpage}--\pageref{lastpage}} \pubyear{2016}

\maketitle

\label{firstpage}

\begin{abstract}
Neutral hydrogen (HI) will soon be the dark matter tracer observed over the largest volumes of Universe thanks to the 21 cm intensity mapping technique. To unveil cosmological information it is indispensable to understand the HI distribution with respect to dark matter. Using a full one-loop derivation of the power spectrum of HI, we show that higher order corrections change the amplitude and shape of the power spectrum on typical cosmological (linear) scales. These effects go beyond the expected dark matter non-linear corrections and include non-linearities in the way the HI signal traces dark matter. We show that, on linear scales at $z = 1$, the HI bias drops by up to 15 \% in both real and redshift space, which results in underpredicting the mass of the halos in which HI lies. Non-linear corrections give rise to a significant scale dependence when redshift space distortions arise, in particular on the scale range of the baryonic acoustic oscillations (BAO).
There is a factor of 5 difference between the linear and full HI power spectra over the whole BAO scale range, which modifies the ratios between the peaks. This effect will also be seen in other types of survey and it will be essential to take it into account in future experiments in order to match the expectations of precision cosmology.
\end{abstract}

\begin{keywords}
Cosmology : Large Scale Structure of the Universe - Cosmology : theory 
\end{keywords}


\section{Introduction}
Future neutral hydrogen (HI) experiments such as the Square Kilometer Array \citep[SKA,][]{2015arXiv150103989S}, its pathfinder MeerKAT, the Canadian Hydrogen Intensity Mapping Experiment \citep[CHIME,][]{2014SPIE.9145E..4VN}, the Hydrogen Intensity and Real-time Analysis eXperiment \citep[HIRAX,][]{2016arXiv160702059N}, and the Baryon acoustic oscillations In Neutral Gas Observations \citep[BINGO,][]{2013MNRAS.434.1239B} will map the cosmological neutral hydrogen within unprecedented volumes of the Universe thanks to the line intensity mapping (IM) technique. This technique relies on the measurement of the HI integrated intensity from hundreds of galaxies in one single large voxel (3D pixel) instead of detecting individual HI galaxies. The observed volumes will allow unrivalled constraints on cosmology \citep[e.g.][]{2015ApJ...803...21B}. Nevertheless, to achieve the expected levels of accuracy one needs to understand how HI relates to the underlying dark matter distribution. Current HI observations are quite sparse. At $z\sim0$, HI is observed in emission at 21 cm but current detections weaken quickly and vanish at $z>0.1$. At intermediate and higher redshifts, the main tracer of HI is Damped Ly-$\alpha$ systems (DLAs), objects with $N_\HI>10^{20.3}$~cm$^{-2}$ displaying a 21~cm line in absorption in the spectrum of a distant quasar. As they are optically thick, hydrogen into their midst is self-shielded and remains neutral. DLAs are actually thought to host most of the neutral gas within $0<z<5$ \citep{1986ApJS...61..249W,1995ApJ...440..435L,2005ARA&A..43..861W} and hence to contain a significant reservoir of neutral gas for star formation at high redshift. Combining emission and absorption measurements, the redshift evolution of the fraction density of HI \OmegaHI~has been shown to decrease slightly from high to low redshift \citep{2015MNRAS.452..217C,2016MNRAS.456.4488S}. Such a mild but somewhat steady evolution leads to the picture of a balance between consumption and replenishment of the gas reservoir.\\
Notwithstanding, even though measurements are used altogether, 21 cm emission and 21 cm absorption line surveys might not target the same population of objects. This is crucial in order to clarify what is the HI bias. DLAs at high redshift might not belong to the same population than HI galaxies at low redshift. Properties of DLA hosts remain largely unknown, either because the background quasar is several magnitudes brighter or because they are too faint to be detected by current spectrographs \citep[$m>25$,][]{2008ApJ...681..856R,2012ApJ...748..121C}. When it comes to the mass of their host dark matter halos, there seems to be a tension between 21 cm low redshift galaxies and DLAs. There are only a handful of measurements of HI and DLA biases. \citet{2012ApJ...750...38M} measured $b_\HI\sim0.8$ at $z\sim0$ in the ALFALFA survey while \citet{2010Natur.466..463C,2013ApJ...763L..20M}, and \citet{2013MNRAS.434L..46S} measured the product \OmegaHI\bHI~in IM data taken with the Green Bank Telescope at $z\sim0.8$. The latter used the IM data in auto-correlation while the former cross-correlated them with galaxy surveys to circumvent the contamination of foregrounds residuals. \citet{2012JCAP...11..059F} measured $b_\mr{DLA} = 2.17\pm0.2$ at $z \sim 2.3$ in the Baryon Oscillation Spectroscopic Survey. Such a value leads to host dark matter halos of $10^{11.5}$\Msun~as compared to the $10^{9-11}$\Msun~found with  21 cm measurements as well as in simulations \citep{2008MNRAS.390.1349P,2014MNRAS.438..529R}. To reconcile bias measurements, \citet{2016MNRAS.458..781P} argued that there must be a significant change in the properties of HI-bearing systems. \\
The knowledge of the HI bias requires to understand how HI populates dark matter halos. Even though it is widely accepted that HI is within galaxies at $z<5$, today a simple relation between dark matter halo and HI masses (MHIMh) is used, with a certain gas profile when necessary (which is not for the bias). The MHIMh relation is measured in hydrodynamical simulations often assuming a simple power law \citep{2013MNRAS.434.2645D,2016arXiv160302383K,2016MNRAS.456.3553V}, inspired from observations \citep{2010MNRAS.407..567B} or parametrised and fitted on data \citep{2010MNRAS.403..870B,2014MNRAS.440.2313B,2016MNRAS.458..781P,2016arXiv160701021P}. Lately, \citet{2016arXiv160800007P} derived the MHIMh relation using the abundance matching technique where the halo mass function is matched to the HI mass function. Even if these different schemes can strongly differ, they all lead to similar values of the linear HI bias.\\
On non-linear scales, the HI bias has been barely investigated yet, while it contains a wealth of information on cosmology and, above all, on the MHIMh relation. To date, the scale dependence of the HI bias has been measured in two ways : in hydrodynamical simulations \citep{2014JCAP...09..050V} and in N-body simulations where halos are populated with HI through an empirical relation \citep{2016MNRAS.460.4310S,2016JCAP...03..001S}. However these methods suffer from a few limitations. In hydrodynamical simulations, the bias is sensitive to the physical processes that are included in the simulation, and, first and foremost, to the resolution. For instance, several zoom-in simulations will not lead to the same value of the bias at the same common scale. In addition, hydrodynamic simulations can hardly access linear scales and both approaches are computationally heavy. The investigation of the influence of the MHIMh scheme on the HI bias on the full scale range requires a more flexible approach.\\
We use the full one-loop calculation of \citet{2016JCAP...03..061U} and \citet{2016arXiv161104963U} to compute the non-linear power spectrum of HI. It relies on high order HI biases that are computed with the halo model for each MHIMh prescription. We compute the power spectrum and the bias of HI in both real and redshift space and show that non-linear terms have a significant contribution on linear scales. We limit our analysis to $z=1$ which is one of the most targeted redshifts for BAO measurements. \\
This paper is organised as follows. We begin with reviewing the theoretical framework of the HI power spectrum and listing the MHIMh relations we use in Sect. \ref{par:theory}. Second, we compute the power spectrum and bias in real space with which we examine the mass of halos in which HI lie in Sect. \ref{par:PHI_real_space}. Third, we carry a similar analysis in redshift space and discuss its cosmological implications in Sect. \ref{par:PHI_redshift_space}. We conclude in Sect. \ref{par:ccl}\\
Throughout the article, we use the Planck 2014 Cosmology \citep{2014A&A...571A..16P}.
\begin{figure}
\includegraphics[scale=0.65]{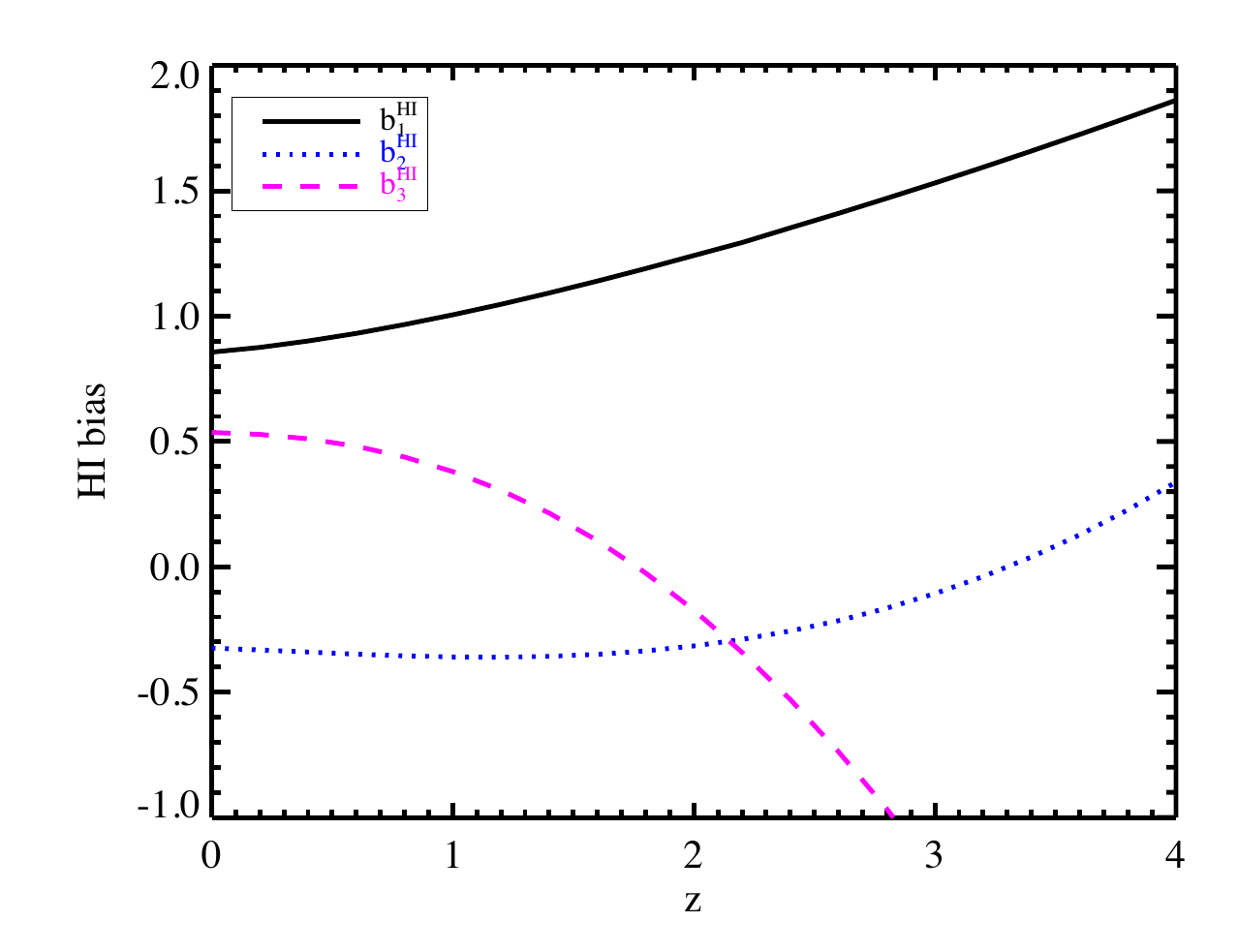} 
\caption{HI biases computed with Eq. \ref{eq:n_order_bias} using the HOD A prescription for the relation between HI and halo mass at $z=1$.}
\label{fig:all_biases}
\end{figure}
\section{Modelling the HI power spectrum}\label{par:theory}
\subsection{The power spectrum}
The average HI brightness temperature is given by \citep{2013MNRAS.434.1239B}
\be 
\overline{T}(z) = 566\, h \left( \frac{H_0}{H(z)} \right) \left(\frac{\Omega_\HI(z)}{0.003}\right) \left(  1 + z \right)^2 \, \, \mu\mr{K}
\label{eq:temperature}
\ee
where the HI density fraction is defined as $\Omega_\HI = \rho_\HI/\rho_{c,0}$ with $\rho_{c,0}$ is the critical density of the Universe today. The fluctuating part is
\be 
T(z,\mathbf{x}) = \overline{T}(z) \left( 1 + \delta_\mr{HI}(\mathbf{x}) \right)
\ee
with $\delta_\mr{HI}(\mathbf{x}) $ the HI density fluctuation at position $\mathbf{x} $, hence, in Fourier space
\be 
\langle T(z,\mathbf{k}) T^\star(z,\mathbf{k'}) \rangle = (2\pi)^3 P_\HI(k,z)\delta^3(\mathbf{k} - \mathbf{k'})
\ee
Carrying a full one-loop derivation of the HI brightness temperature in Perturbation Theory \citep{2002PhR...367....1B}, the power spectrum of HI in real space at redshift $z$  is 
\be 
P_\HI (z,k)  = P_\HI^{11}(z,k)  + P_\HI^{22}(z,k)  + P_\HI^{13} (z,k) 
\ee
where $P_\HI^{11}(z,k)$ is the linear power spectrum (tree level) while $P_\HI^{22}(z,k)$ and $P_\HI^{13}(z,k)$ are the non-linear corrections. For clarity purposes we will not specify the redshift dependence in the following. Following \citet{2016JCAP...03..061U} and \citet{2016arXiv161104963U} the three terms of $P_\HI (k)$ are 
\bea
P_\HI^{11}(k) &=& \overline T^2\, b_1^2\, P_\m^{11}(k)\\
P_\HI^{22}(k) &=& \frac{\overline T^2}{2}\int \frac{\dd^3 k_1}{(2\pi^3)} \Big[ b_1\, F_2(\k_1,\k_2) +  b_2 \Big]^2 \nonumber\\
&\times& P_\m^{11}(k_2)\, P_\m^{11}(k_1)\\
P_\HI^{13}(k) &=& \overline T^2\,b_1 \Big\{ \left(b_3 + \frac{68}{21}b_2\right)\sigma_\Lambda^2  P_\m^{11}(k) \nonumber\\
&+& b_1\, P^{13}_\m(k)\Big\}
\eea
where $k_2 = |\k_1 - \k|$. $b_1, b_2$, and $b_3$ are the linear, second and third order HI biases, respectively. The latter are the higher terms of the bias expanded in Taylor series, which means assuming that the HI bias is local. $F_2$ is the non-linear density kernel defined in Appendix \ref{app:ir_int}. Finally $\sigma_\Lambda$, the variance of the dark matter field, is 
\be 
\sigma_\Lambda^2 = \int_{k_\mathrm{min}}^{k_\mathrm{max}}\frac{\dd^3 k}{(2\pi)^3} P_\m(k)
\ee
For simplicity, we set ${k_\mathrm{max}}$ to the non-linear dispersion scale, $k_\mathrm{NL} = 0.2\, h (1+z)^{2/(2+n_s)}$ Mpc$^{-1}$ with $n_s$ the spectral index. \\
In redshift space, the 3D power spectrum of HI on linear and quasi-linear scales at scale $k$, and $\mu$, the cosine of the angle between the line of sight and the separation vector $\k$, writes 
\be 
P_\HI (k,\mu)  = P_\HI^{11}(k,\mu)  + P_\HI^{22}(k,\mu)  + P_\HI^{13} (k,\mu) 
\ee
Following \citet{1987MNRAS.227....1K}, the linear term in redshift space is 
\be 
P_\HI^{11}(k,\mu) = \overline T^2 \left[b_1 + f\, \mu^2\right]^2\, P_\m^{11}(k)
\label{eq:PHI_11}
\ee
with $\mu = k_\parallel/k$, $P_\m(k)$ the linear power spectrum of matter, and $f$ the linear growth rate. We compute the former using the transfer function of \citet{1998ApJ...496..605E} and assume $f(z) = \Omega_\m(z)^\gamma$ with $\gamma = 0.55$ for $\Lambda$CDM \citep{1980lssu.book.....P,2005PhRvD..72d3529L}.\\
Following \citet{2016JCAP...03..061U} and \citet{2016arXiv161104963U} the one-loop corrections are 
\bea 
P_\HI^{22}(k,\mu) &=& \frac{\overline T^2}{2}\int \frac{\dd^3 k_1}{(2\pi)^3} \Big[ b_1\, F_2(\k_1,\k_2) \nonumber \\
 &+& \mu^2 G_2(\k_1,\k_2) + b_2 + K_R(\k_1,\k_2)  \Big]^2 \nonumber \\
&\times& P_\m^{11}(k_2)\, P_\m^{11}(k_1)\\
P_\HI^{13}(k,\mu) &=& \overline T^2 \left(b_1 + \mu^2 f\right) \nonumber\\
&\times&\Big\{  \left[  \left(b_3 + \frac{68}{21}b_2\right)\sigma_\Lambda^2 +I_R(k,\mu)  \right] P_\m^{11}(k) \nonumber\\
&+& \left[b_1\, P^{13}_\m(k) + \mu^2\,f\, P^{13}_\theta(k) \right]\Big\}
\label{eq:PNL_redshift_space}
\eea
where $P^{13}_\m(k)$ and $P^{13}_\theta(k)$ are the third order matter power spectrum and velocity field power spectrum, respectively. Their expressions along with that of $I_R(k,\mu)$ are listed in Appendix \ref{app:ir_int}. Finally, several kernels are involved in the computation of the $P_\HI^{22}(k,\mu) $ term : $G_2$ induced by peculiar velocities at second order and $K_R$ arises from non-linear mode coupling  \citep[velocity-velocity and velocity-density,][]{2002PhR...367....1B}. Their expressions are also given in Appendix \ref{app:ir_int}.
\begin{figure}
\includegraphics[scale=0.65]{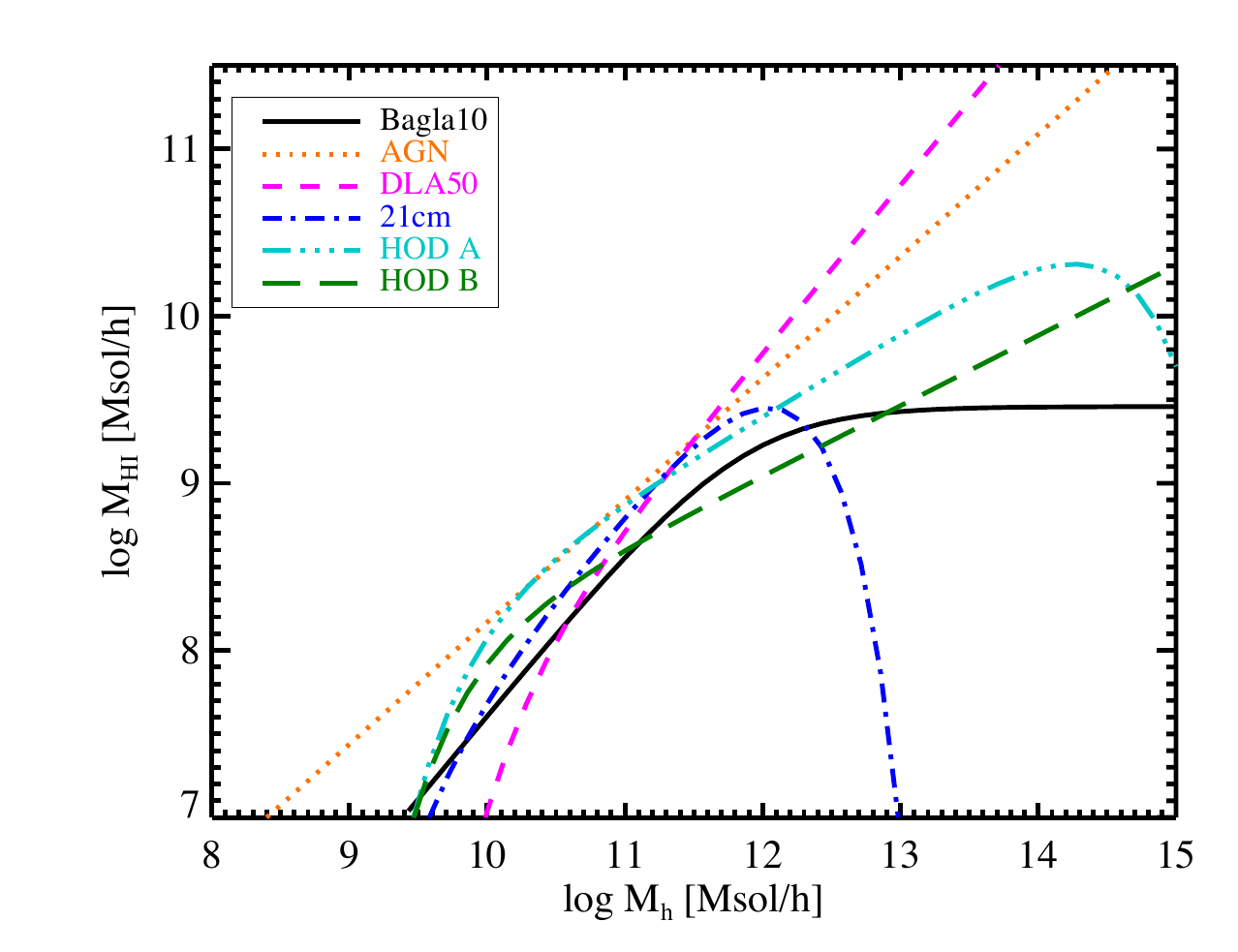} 
\caption{Relations between the HI mass and the halo mass at $z = 1$.}
\label{fig:MHI_of_M_vs_models}
\end{figure}
\subsection{HI quantities}
HI-related quantities such as densities and biases are computed using the halo model that provides a description of the clustering of dark matter halos at both linear and non-linear scales  \citep{2002PhR...372....1C}. It relies on the halo mass function $\dd n/ \dd M$ and the associated $n-$th order halo biases $b_n^h(M)$ measured in N-body simulations. We use the prescriptions of  \citet{1999MNRAS.308..119S}. \\
The comoving density of HI writes 
\be 
\rho_\HI = \int \dd M \frac{\dd n}{\dd M} \, M_\HI(M)
\ee
The $n$-th order HI biases are
\be 
b_n^\mathrm{HI} = \frac{1}{\rho_\HI}\int \dd M \frac{\dd n}{\dd M} b_n^\mr{h}(M)\, M_\HI(M)
\label{eq:n_order_bias}
\ee
where $M_\HI(M)$ is the relation between the HI mass and the halo mass (see Sect. \ref{par:MHI_MH_def}).
Fig. \ref{fig:all_biases} shows an example of a set of biases. Note that only the first order bias is always positive while the two others change sign. All of them increase for high halo masses. We will use the terms \textit{linear} and \textit{first order} bias interchangeably.
\subsection{The HI mass - halo mass relation}\label{par:MHI_MH_def}
The distribution of HI within the Large Scale Structure is rather unclear today. It is believed that in the post-reionization era most of HI lies within galaxies while only a negligible fraction is diffuse \citep{2016JCAP...03..001S}. It is often simply parametrised by relating the mass of HI to the mass of its host dark matter halo through a simple power law including, or not, a cut-off at small and high halo masses. We compile here several MHIMh relations that have been used or estimated using both hydrodynamical simulations and parametrised models fitted on data measurements. We also consider a DLA model.
\begin{table*}\centering 
\begin{tabular}{ccccccc}
\hline\hline 
Model                 & Parameters                                                                                                              & $b_1$             & $b_2$      & $b_3$   & $b_\mathrm{eff}$  & $T_\HI \times 10^{4}$ K  \\
\hline
Bagla10              &    None                                                                                                                    &  0.93              & -0.41      &   0.62     & 0.80                   &  1.65\\
AGN                   & $\alpha = 0.73$, $\gamma = 2$                                                                            & 0.91               &-0.27       &  0.41      & 0.82                   & 12.14\\
21cm                 & $\alpha$ = 0.15                                                                                                      &  0.96              &-0.42       &   0.60     & 0.81                   &  2.43\\
HOD A               & $\log v_{c,0}$ = 1.58 ,  $\log v_{c,1}$ = 3.14, $\alpha =0.17$ , $\beta = -0.5$       & 1.00               & -0.35      &  0.38       & 0.82                  &  4.38 \\
HOD B               & $\log_{10}v_{c,0}$ = 1.56 , $\alpha =0.09$ , $\beta = -0.58$                                    & 0.96               & -0.37      &  0.49      & 0.85                   & 2.55\\
\hline
DLA50              &  $\alpha = 0.13$                                                                                                       &  1.64              &  0.56      & -1.27      & 1.74                  & 49.94\\
\hline 
\end{tabular}
\caption{Free parameters of the MHIMh prescriptions along with the associated HI biases and mean temperatures at $z=1$.}
\label{tab:free_parameters+biases}
\end{table*}
\begin{enumerate}
\item \label{num:Bagla10} \textbf{Bagla10} : One relation that has been widely used is that of \citet{2010MNRAS.407..567B}. It has been inspired from quasar observations and assumes that there is no HI in high mass halos  : 
\be 
M_\HI(M) = \frac{f_3\, M}{1+\frac{M}{M_\mr{max}}} \;\;\; \mbox{for} M\geq M_\mr{min}
\ee
where $f_3$ comes from the normalisation to \OmegaHI. This prescription is commonly used for studies of 21 cm intensity mapping \citep[amongst others,][]{2014JCAP...09..050V,2016MNRAS.460.4310S,2016JCAP...03..001S}. \Mmin~and \Mmax~are the limits for a dark matter halo to host HI. They assume that only halos with 30 km/s $<v_\mr{circ}<$200 km/s host HI, which translates to lower and upper bounds,  \Mmin~and \Mmax, through
\be 
v_\mathrm{circ} = 30 \sqrt{1+z} \left( \frac{M}{10^{10} M_\odot}\right)^{1/3} \,\, \mbox{km/s}
\label{eq:vcirc_M}
\ee
\item \textbf{AGN} : Nevertheless, \citet{2016MNRAS.456.3553V} measured the MHIMh relation in hydrodynamical simulations including AGN feedback and show that there is HI in halos that have $v_\mr{circ}>$200 km/s. They measured $M_\HI(M) = e^\alpha\, M^\gamma$ and fit $\alpha$ and $\gamma$ up to redshift 2.
\item \label{num:DLA50} \textbf{DLA50} : A prescription adapted from DLA studies \citep{2010MNRAS.403..870B,2014MNRAS.440.2313B} by \citet{2016MNRAS.458..781P}
\be 
M_\HI(M) = \alpha\,  f_\mathrm{H,c}\, M\, \exp\left[-\left( \frac{v_{c,0}}{v_c(M)} \right)^3 \right] \exp\left[-\left( \frac{v_{c,1}}{v_c(M)} \right)^3 \right]
\label{eq:MHI-MH_padmanabhan}
\ee
where $\alpha$ is the ratio of HI within halos and cosmic HI, $ f_\mathrm{H,c} = (1-Y_p)\Omega_b/\Omega_m$ is the cosmic hydrogen fraction with $Y_p$ the cosmological helium fraction by mass, and  $v_c(M)$ is the virial velocity of a halo \citep{2001MNRAS.321..559B} : 
\be 
v_c(M) = 96.6 \, \mbox{km/s} \left( \frac{\Delta_v\Omega_mh^2}{24.4}\right)^{1/6} \left( \frac{1+z}{3.3}\right)^{1/2} \left( \frac{M}{10^{11}M_\odot} \right)^{1/3}
\ee
with $\Delta_v$ the mean overdensity of the halo that we take to be 200. For DLAs, \citet{2016MNRAS.458..781P} considered $v_{c,0} = 50$~km/s and an infinite $v_{c,1}$. They fitted $\alpha$ to measurements between redshift 0 and 4  (column density distributions, biases, \OmegaHI and the incidence rate). 
\item \label{num:21cm} \textbf{21cm} :  \citet{2016MNRAS.458..781P} adapted Eq.~\ref{eq:MHI-MH_padmanabhan} to 21 cm IM observations using ad-hoc velocity cuts $v_{c,0}=30$~km/s and $v_{c,1}=200$~km/s. Similarly to the DLA50 model, \citet{2016MNRAS.458..781P} fitted $\alpha$ on the same measurements. Note that in both latter cases the slope is fixed and equal to unity which is higher than what is measured in hydro-simulations.
\item  \label{num:HODA} \textbf{HOD A} : \citet{2016arXiv160701021P} improved Eq. \ref{eq:MHI-MH_padmanabhan} by introducing a flexible slope, $\beta$, as well as the velocity cut-offs:
\bea 
M_\HI(M) &=& \alpha\,  f_\mathrm{H,c}\, M\,  \left(  \frac{M}{10^{11} h^{-1} M_\odot}   \right) ^\beta \exp\left[-\left( \frac{v_{c,0}}{v_c(M)} \right)^3 \right] \nonumber\\
&\times&\exp\left[-\left( \frac{v_{c,1}}{v_c(M)} \right)^3 \right]
\label{eq:MHI-MH_padmanabhan3}
\eea
where $\alpha$, $\beta$, $v_{c,0}$, and $v_{c,1}$ are free parameters and fitted on data measurements.
\item  \label{num:HODB} \textbf{HOD B} : Lastly, \citet{2016arXiv161106235P} fitted an updated version of Eq. \ref{eq:MHI-MH_padmanabhan3} 
\be 
M_\HI(M) = \alpha\,  f_\mathrm{H,c}\, M\,  \left(  \frac{M}{10^{11} h^{-1} M_\odot}   \right) ^\beta \exp\left[-\left( \frac{v_{c,0}}{v_c(M)} \right)^3 \right]
\label{eq:MHI-MH_padmanabhan2}
\ee
on all the available measurements including galaxy clustering. Their free parameters are $\beta$ and $\alpha$.
\end{enumerate}
All these prescriptions are shown in Fig. \ref{fig:MHI_of_M_vs_models} at $z=1$. They vary in shape, amplitude, and slope. Clearly the DLA50 scheme favours high halo masses as compared to the other models. We limit our analysis to $z=1$, the values of the free parameters are given in Table \ref{tab:free_parameters+biases}.
\begin{figure*}
\includegraphics[scale=0.7]{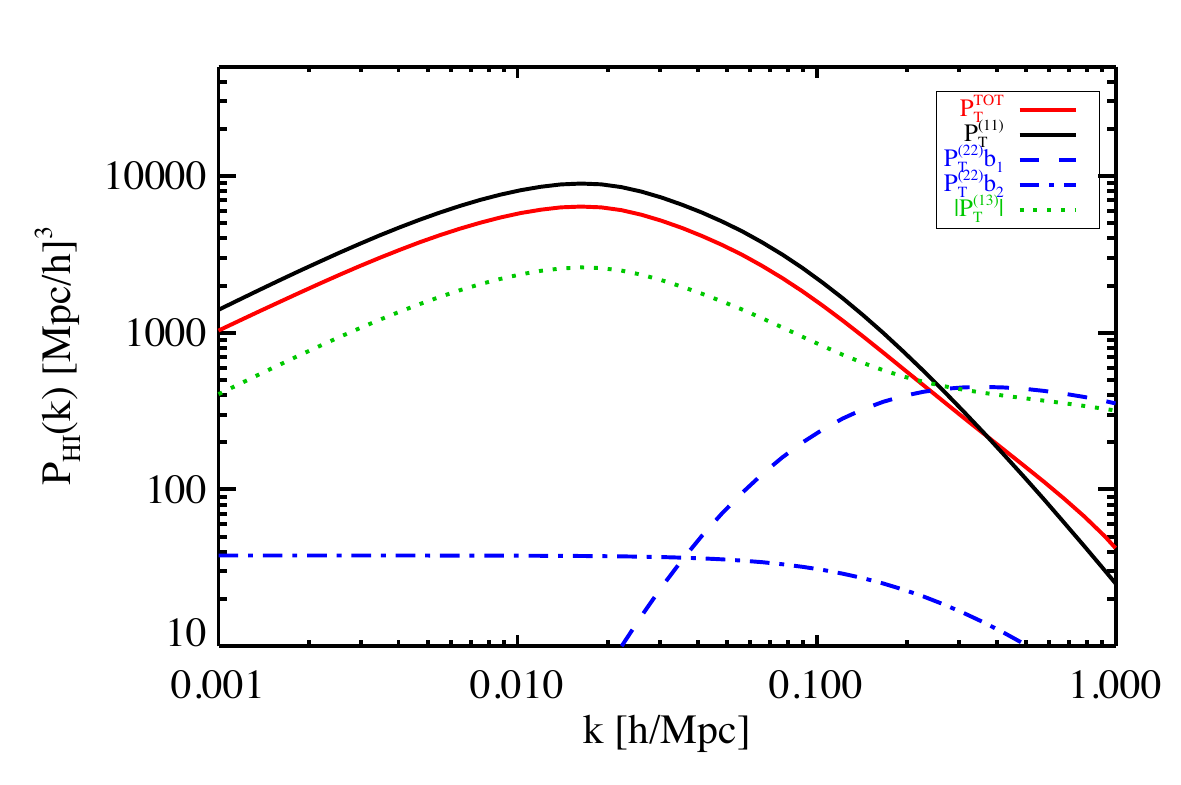} \includegraphics[scale=0.7]{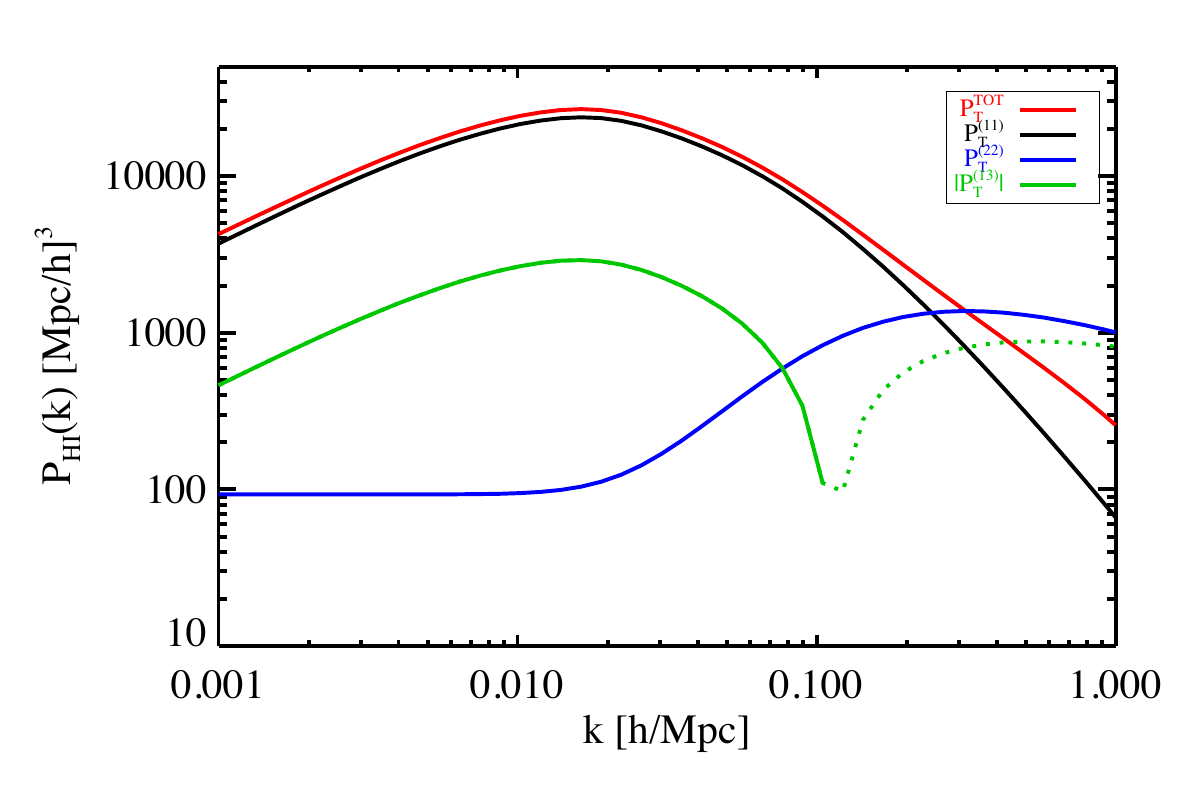} 
\caption{Linear and non-linear contributions to the real space HI power spectra for the HOD A prescription (left) and the DLA50 one (right). Note that green lines show the absolute value of the $P_{13}$ term and the dotted line is the negative part. The dashed and dot-dash blue lines are the $b_1$ and $b_2$ components of the $P_{22}$ term, respectively.}
\label{fig:PHI_all_terms_real_space}
\end{figure*}
\section{The HI power spectrum in real space}\label{par:PHI_real_space}
In this section, we compute the non-linear HI power spectrum in real space using all the above MHIMh models. We first describe the non-linear contributions to the power spectrum and show that the bias is neither constant nor linear on so-called linear scales. We discuss the implications of that effective bias on our understanding of the distribution of HI and compare this modeling approach to others.
\subsection{A non linear bias on linear scales}
\begin{figure*}
\includegraphics[scale=0.7]{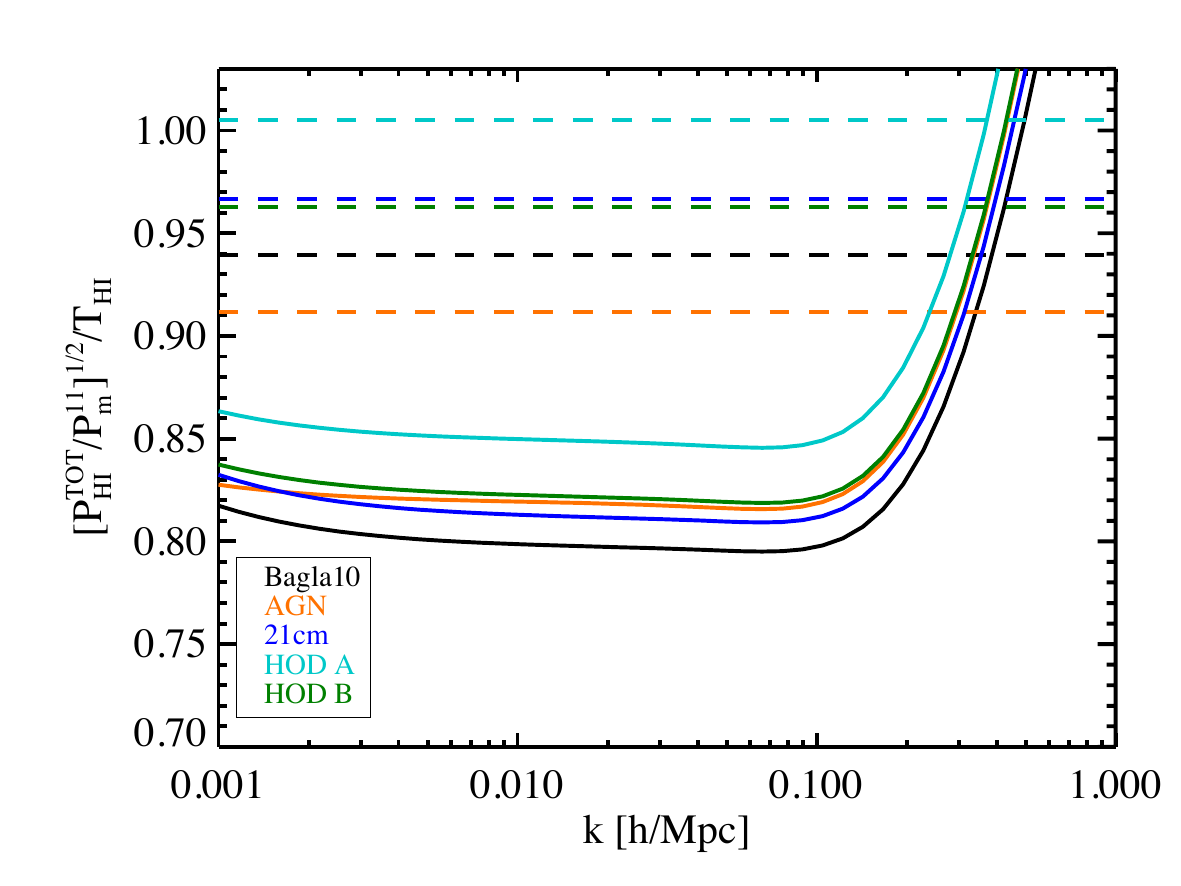} 
\includegraphics[scale=0.7]{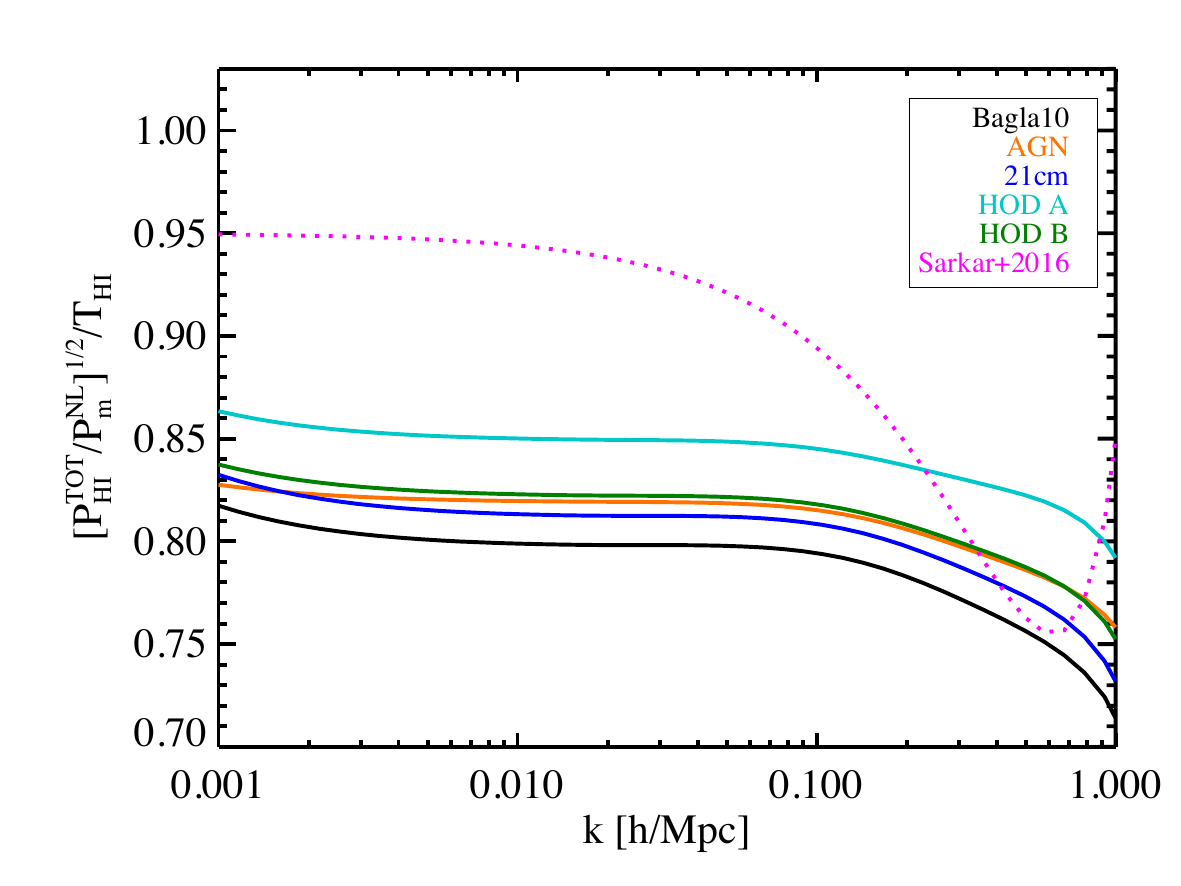} 
\caption{Scale dependence of the HI bias (left) and the ratio $\sqrt{P_\mathrm{HI}(k)/P_\mathrm{m}^\mathrm{NL}(k)}/\overline T_\mathrm{HI}$ (right) for the different MHIMh prescriptions in real space. Horizontal dashed lines are the linear biases for each MHIMh models computed with Eq. \ref{eq:n_order_bias}. The magenta dotted line is the HI bias of \citet{2016MNRAS.460.4310S}. The $P_\mr{m}^{11}$ and $P_\mr{m}^\mr{NL}$ are the linear and non-linear power spectra of matter, respectively.}
\label{fig:biasHI_vs_all_models}
\end{figure*}

The total HI power spectrum along with the non-linear contributions are shown in Fig. \ref{fig:PHI_all_terms_real_space} for model HOD A. Contrary to our expectations, both $P_\HI^{22}$ and $P_\HI^{13}$ terms have significant contributions on linear scales. These contributions arise from the coupling of short and long wavelength modes. The $P_\HI^{13}$ term is negative and proportional to the matter power spectrum. Therefore, on linear scales, it lowers the amplitude of the HI power spectrum by $\sim25$ \% as compared to a standard biased power spectrum. Hence, the actual HI bias is lower than the linear HI bias. The $P_\HI^{22}$ term is constant on linear scales which induces a scale dependence of the HI bias on the largest scales. The flat contribution to the $P_\HI^{22}$ term (the dot-dashed line) is simply proportional to $b_2^\HI$ while, on those scales, the $P_\HI^{13}$ term is a function of $b_1^\mathrm{HI}, b_2^\mathrm{HI}$, and $b_3^\mathrm{HI}$. The latter depend on the MHIMh prescriptions and are listed in Table~\ref{tab:free_parameters+biases}. While the $b_1^\mathrm{HI}$s vary by $\sim13\%$ amongst the different prescriptions, the variation strongly increases at higher orders. Indeed $b_2^\mathrm{HI}$s and $b_3^\mathrm{HI}$s differ by $35\%$ and $74\%$, respectively. Hence, the shape of the HI bias depends on the MHIMh prescription as shown in Fig. \ref{fig:biasHI_vs_all_models}. Hereafter, we will callHI \textit{effective} bias the following : 
\be 
b_\mathrm{HI}^\mr{eff}(k) = \frac{1}{\overline T_\mathrm{HI}}\sqrt{\frac{P_\mathrm{HI}(k)}{P_\mathrm{m}^\mathrm{11}(k)}} 
\label{eq:computed_bias}
\ee
The right panel of Fig. \ref{fig:biasHI_vs_all_models} shows the ratio $1/\overline T_\mathrm{HI}\sqrt{P_\mathrm{HI}(k)/P_\mathrm{m}^\mathrm{NL}(k)}$ where $P_\mathrm{m}^\mathrm{NL}(k)$ is the non-linear matter power spectrum computed using the same perturbation theory framework (See Appendix \ref{app:ir_int}). The normalisation by the mean HI temperature is to focus only on the bias and avoid additional amplitude variations. Indeed, the HI temperature is a function of $\Omega_\HI$, therefore of the MHIMh relation (see Eq. \ref{eq:temperature} and Table \ref{tab:free_parameters+biases}). On the largest scales, there is only a few percent difference between the different prescriptions. We won't discuss them here as General Relativity corrections must be taken into account on ultra large scales. On non-linear scales, regardless of the MHIMh relation, we recover a bias well below 1, meaning that HI galaxies are highly anti-biased while they are only slightly on linear scales \citep{2010ApJ...718..972M, 2012ApJ...750...38M}. At $k>0.2\,h\, $Mpc$^{-1}$ our biases are of the same order of magnitude than that of \citet{2016MNRAS.460.4310S} as shown in the right panel of Fig. \ref{fig:biasHI_vs_all_models}. Their dip is deeper because their bias on linear scales is higher than ours. Nevertheless, they are consistent as explained in the following and in Sect. \ref{par:modeling}. \\
On linear scales, there is, at most, a difference of 10\% in the amplitude of the models and 15\% at $k=1\, h$ Mpc$^{-1}$. Regardless of the MHIMh prescription, the \textit{effective} bias $b_\HI^\mathrm{eff}$, is lower than the linear one. Their values at $k = 0.01\, h$~Mpc$^{-1}$ are listed in Table~\ref{tab:free_parameters+biases} along with the associated linear biases. On linear scales, effective biases are always lower by 10-15\% than their linear counterpart. They can be approximated, in real space, by 
\be 
b_1^\HI \rightarrow b_\HI^\mathrm{eff} \approx b_1^\HI + \frac{1}{2}\left(  b_3^\HI + \frac{68}{21}b_2^\HI \right) \sigma_\Lambda^2
\label{eq:bias_eff}
\ee
It has consequences on our understanding of HI within the Large Scale Structure. The assumption that the measured HI bias on large scales is linear leads to an underestimation of that linear bias and therefore of the halo mass hosting HI. Hence HI lies in slightly more massive halos than thought.
%
%
\subsection{In which halos does HI lie?}
Currently, there is a tension between halo masses of HI-bearing systems, in particular between observations of HI galaxies at low redshift and those of DLAs at higher redshift as highlighted by \citet{2016MNRAS.458..781P}. They fitted all HI available measurements (DLA incidence rates, column densities, biases, HI fractional densities and biases) at several redshifts with both DLA- and 21 cm-based models (our schemes DLA50 and 21cm, amongst others). They showed that the 21 cm based model fitting all measurements systematically underpredicts the DLA bias. Similarly, DLA models, that are tuned to reproduce the DLA bias, always overpredict \OmegaHI\bHI. Their two DLA models have low velocity cut-offs of 50 and 90 km/s, which implies that there is no or only a low amount of neutral hydrogen in low mass halos. While \citet{2014MNRAS.440.2313B} suggested that it could be caused by a strong stellar feedback, it remains inconsistent with HI observations at low redshift. In addition, the discrepancy holds when varying the HI concentration. It is important to note that the discrepancy is only at the level of the biases, hence, the tension is between the host halos of low and high redshift HI-bearing systems. \citet{2016MNRAS.458..781P} argued that there must be a dramatic change in the properties of these systems over $0<z<3$ to have these halo masses on the same evolution path. This idea is strengthened by \citet{2016arXiv160701021P} who introduced a halo occupation model inspired by both DLA and 21 cm emission framework (our HOD A model), using the same dataset as the former together with the HI mass function at $z\sim0$. Again, most of the observables are relatively well fitted but the DLA bias is, again, underpredicted while the high mass part of the HI mass function is overpredicted. Lately, \citet{2016arXiv161106235P} carried a similar analysis with an updated version of the MHIMh relation (our HOD B model) adding the 2-point correlation function (2PCF) of HI galaxies at small scales. By levering some degrees of freedom in the HI concentration, they did improve the overall quality of the fit but with an overpredicted 2PCF on large scales, a high mass tail of the HI mass function, and a too low DLA bias. It is clear that both models predict too many objects in high mass halos and a HI bias that is too high. Indeed, fits are driven towards high halo masses by the DLA bias measurement, which might be flawed as it is inconsistent with most observations and simulations. The latter statement seems inconsistent with our previous argument which is that HI lies in more massive halos than we think. The non-linear corrections to the bias are of the order of 15\% at most while the discrepencies between the predicted and measured HI biases are, at least, of 50\%. Therefore the systematic error due to the assumption of a linear bias on large scales is concealed by the error induced by the DLA bias. \\
We also consider a MHIMh relation adapted to DLAs, the DLA50 model. It is obvious from Fig. \ref{fig:MHI_of_M_vs_models} that it favours higher mass halos as compared to any other prescription. This translates to a higher linear bias and to a change of sign of $b_2^\mathrm{HI}$ and $b_3^\mathrm{HI}$(see Table \ref{tab:free_parameters+biases}). $P_\HI^{13}$ becomes positive on large scales as shown in Fig. \ref{fig:PHI_all_terms_real_space} which adds power to the HI power spectrum. Hence, the effective bias is higher than its linear counterpart in the case of DLAs. The amplitude of the power spectrum rises by 13\% which translates in 7\% on the effective bias for the DLA50 model. Of course, the additional power increases when going towards even higher mass halos. For instance, using a $v_{c,0}$ = 90 km/s instead of $v_{c,0}$ = 50 km/s, which translates to minimum halo masses of $10^{10.04}$ and  $10^{10.80}\, h^{-1}\, M_\odot$ at $z=1$, leads to an increase of power of 24\% and 11\% at the power spectrum and bias levels, respectively. Thus, DLA models overpredict the HI bias even more than previously thought which enhances the tension between DLAs and 21 cm biases preventing any reconciliation. 
%
%
%
%
\subsection{Consistency with other modeling approaches and clustering analysis}\label{par:modeling}
It is the coupling between small and large scale modes that gives rise to an effective bias different from the linear one. Therefore, the mismatch exists for any tracer of dark matter. Hence one can wonder why it is not predicted by any other modeling approaches and why it has never been noticed in any clustering analysis. The answer to the first question is straightforward~: models are constructed to predict a linear bias on linear scales. The procedure for modeling the clustering of any tracer is a distribution of halos coming either from the halo model or dark matter simulations which are filled in with the tracer. Therefore, only non-linearities coming from the evolution of the distribution of dark matter are present and not the ones coming from the distribution of the tracer, which is not the case in our approach. Indeed, it is the distribution of the tracer, the HI brightness temperature precisely, that has been perturbed. Therefore, our biases in Fig. \ref{fig:biasHI_vs_all_models} are consistent with that of \citet{2016MNRAS.460.4310S}. They used a dark matter simulation in which they defined halos that are assigned an HI mass through the Bagla10 model. On the largest scales, they measure a HI bias of 0.92 fully consistent with the linear bias of 0.94 computed through Eq. \ref{eq:n_order_bias}. \\
Lastly, why a mismatch has not been noticed in clustering analysis yet as linearities are missing in current models? On linear scales, it is an offset of 15\% on the bias for HI at most and the scale dependency is only of a few percent so it can be well concealed in the error bars. For instance, when fitting the parameters of a halo occupation distribution on a correlation function on both linear and non-linear scales, the halo mass thresholds hosting a central galaxy and one satellite galaxy would be found to be lower than it is in reality. Notwithstanding, this systematic error is lower than the statistical error on the fitted parameters in current clustering analysis. It won't be the case with stage IV experiments such as Euclid and SKA. In addition, in the era of precision cosmology, ignoring these corrections will lead to flawed estimations of cosmological parameters. 
\begin{figure*}
\includegraphics[scale=0.65]{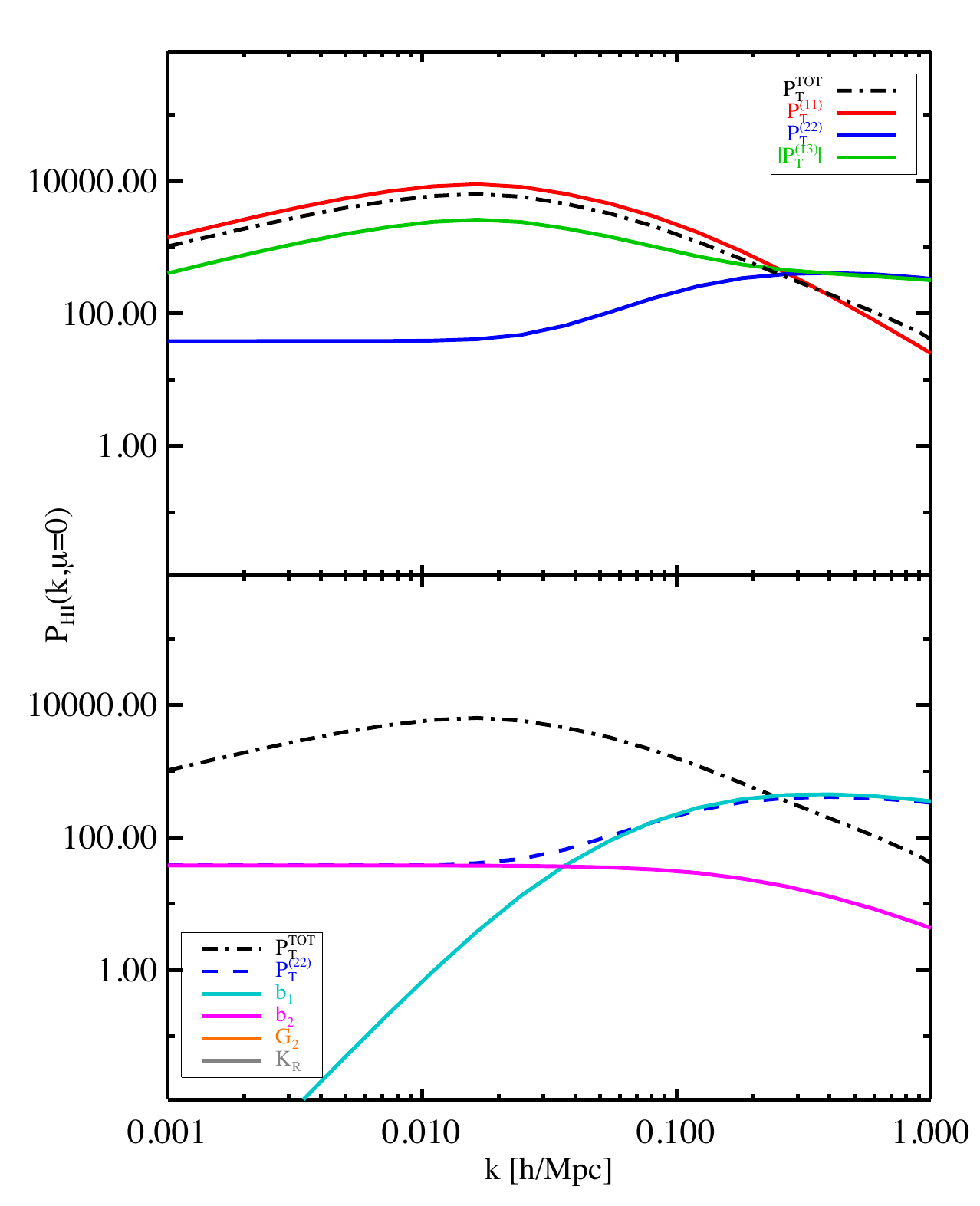}  \includegraphics[scale=0.65]{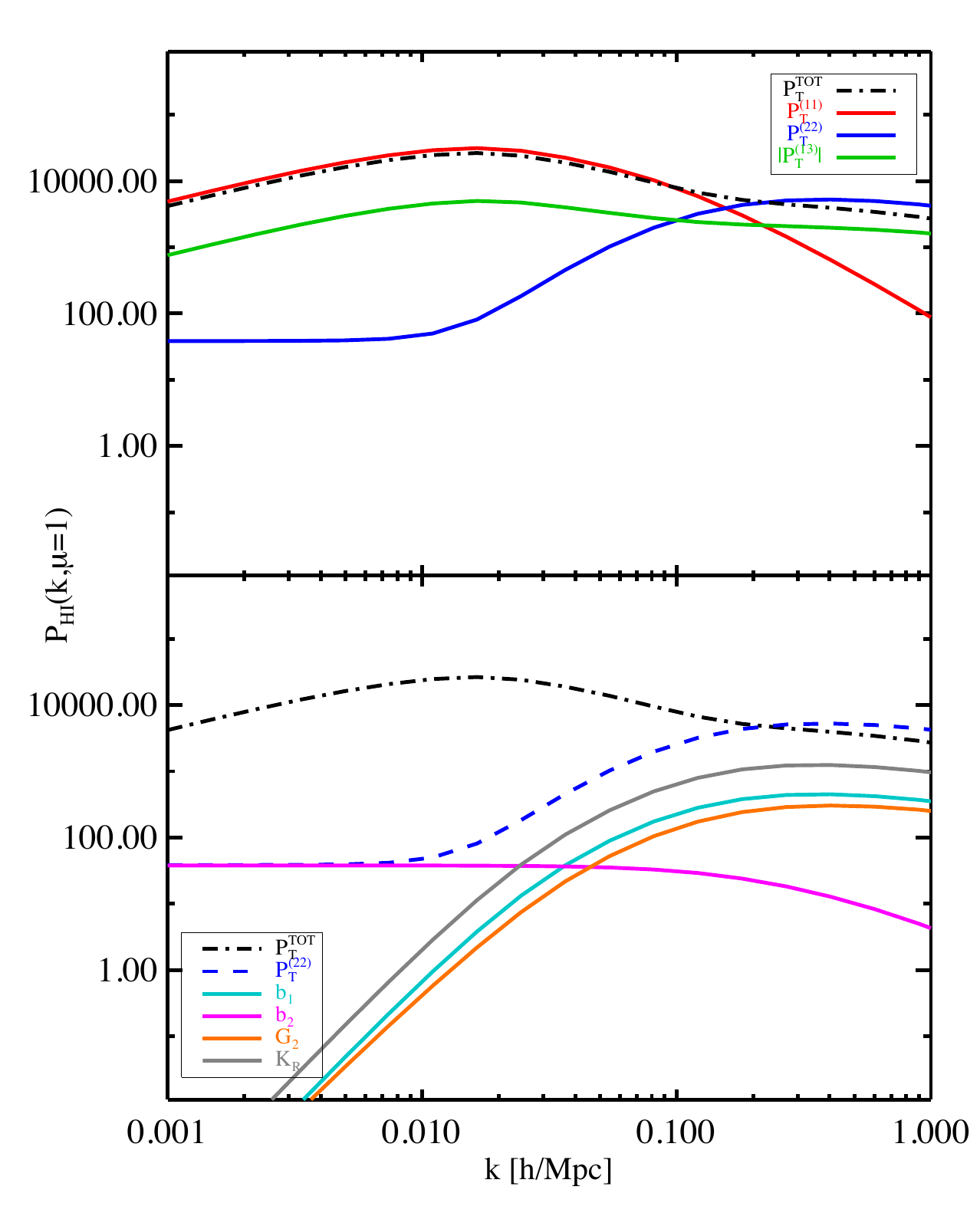} 
\caption{Anisotropic power spectrum in the transverse (left) and radial (right) directions for the MHIMh model HOD A. Top panels show the different non-linear terms while bottom panels show the detailed contributions to the $P_\HI^{22}$ term.}
\label{fig:PHI_terms_z_space}
\end{figure*}

\section{The HI power spectrum in redshift space}\label{par:PHI_redshift_space}
In this section, we extend the previous analysis to the anisotropic power spectrum of HI in redshift space. We first adopt a theoretical point of view, investigating the power spectrum as a function of $k$ and $\mu$ to understand the effects of RSDs and second, we compare the expected linear power spectrum to the full one in the transverse and radial directions.
\begin{figure}
\begin{tabular}{c}
\includegraphics[scale=0.7]{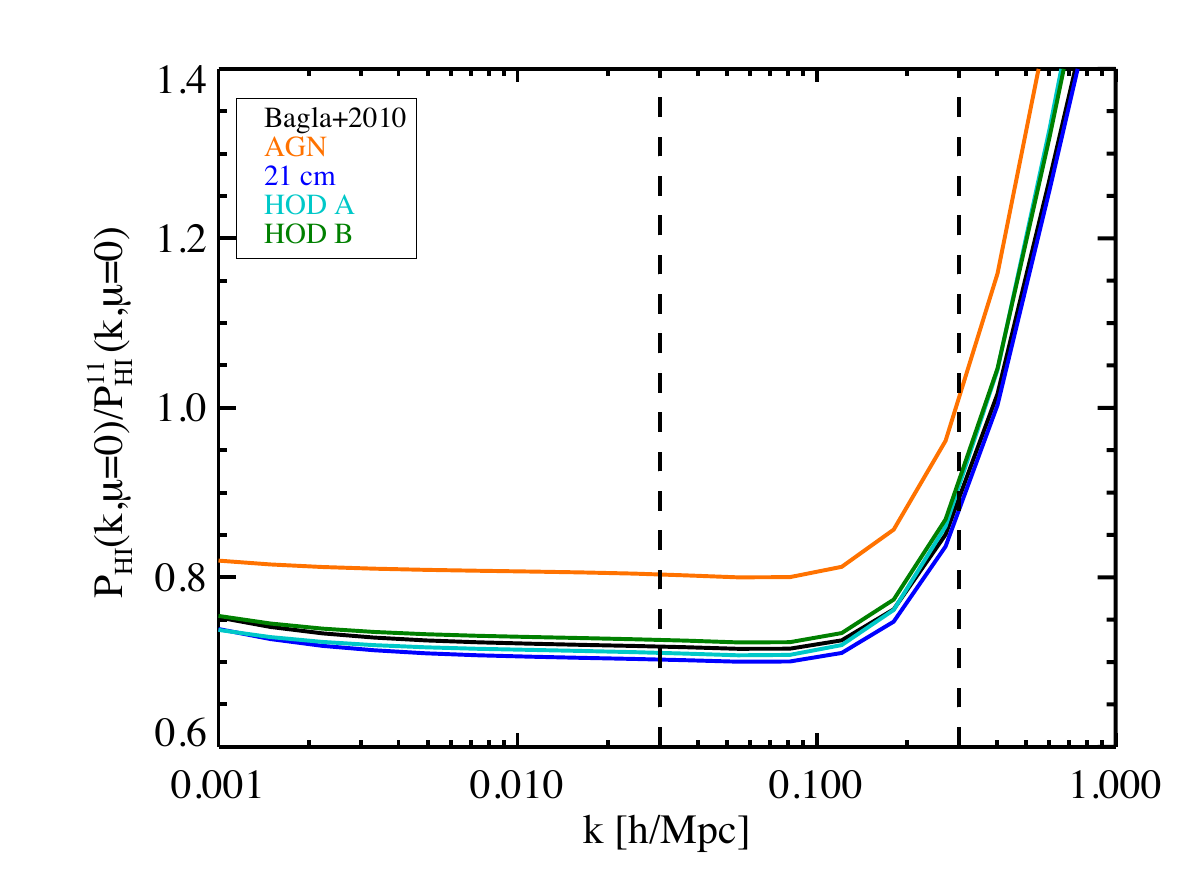}\\
\includegraphics[scale=0.7]{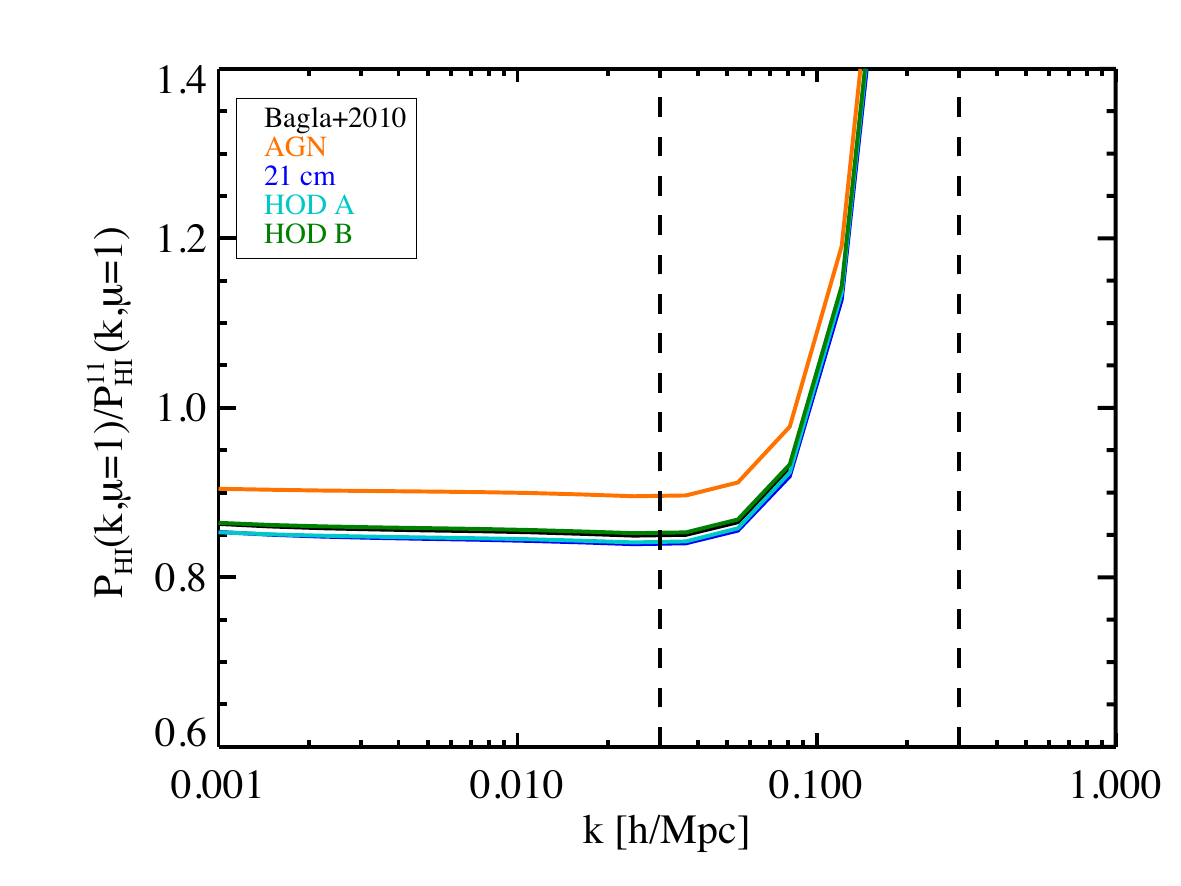}\\
\includegraphics[scale=0.7]{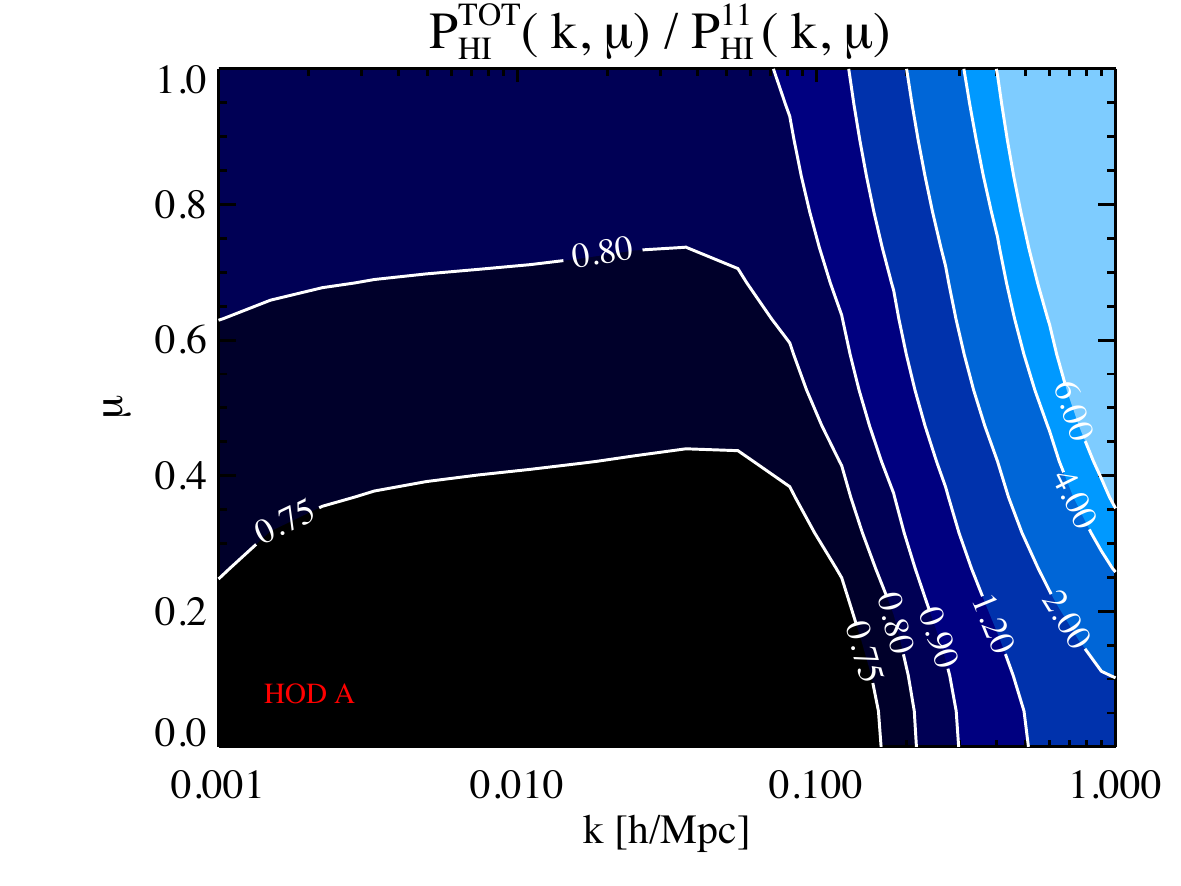}\\ 
\end{tabular}
\caption{Ratios of the total HI power spectrum with the linear HI power spectrum for all the MHIMh prescriptions in the transverse (top panel) and the radial (middle panel) directions at $z=1$. The lower panel shows the same ratio as a function of both $\mu$ and $k$ for HOD A model. Black dashed lines are the BAO scale limits.}
\label{fig:ratio_PHI_P11}
\end{figure}
\subsection{The HI effective bias on linear scales}
We begin by investigating the different contributions to the HI power spectrum in the two extreme directions: in the transverse one where $\mu = 0$, meaning that RSDs are null, and in the radial direction where $\mu=1$ and, hence, RSDs are maximal. Fig. \ref{fig:PHI_terms_z_space} shows the different terms contributing to the HI power spectrum for the prescription HOD A. We recover similar behaviours to those in real space. The $P_\mathrm{HI}^{22}$ term is constant on large scales ($k<0.02\,h\, $Mpc$^{-1}$) and rises towards small scales. The amplitude of the rise increases with $\mu$ as RSDs come in, they are contained in the $G_2$ and $K_R$ terms as shown in the lower panel. Therefore, on small scales ($k>0.2\,h\, $Mpc$^{-1}$) non-linear contributions are maximal for $\mu=1$ where fingers of God are recovered. Again, the $P_\mathrm{HI}^{13}$ term is negative and thus removes power to the HI power spectrum on linear scales. The amplitude of the removal decreases with $\mu$:  at the power spectrum level it lowers from $\sim25\%$ at $\mu = 0$ to $\sim 13\%$ at $\mu = 1$, respectively. The former is similar to the real space case. The effective bias in redshift space is
\be 
b_1^\HI \rightarrow b_\HI^\mr{eff}(k,\mu) \approx  b_1^\HI + \frac{1}{2}\left[\left(   b_3^\HI + \frac{68}{21}b_2^\HI \right) \sigma_\Lambda^2 + I_R(k,\mu)\right]
\ee
The expression of $I_R$ is given in Appendix \ref{app:ir_int} and it is worth noticing that the effective bias is also a function of the growth factor. We will explore this in more details in future work.\\
For $\mu\neq 0$ the effective bias cannot be computed directly because of RSD effects so we compare the full HI power spectrum to the linear Kaiser prediction $P_\HI^\mr{11}(k,\mu) = \overline T^2 \left[b_1 + f\, \mu^2\right]^2\, P_\m^{11}(k)$ in Fig. \ref{fig:ratio_PHI_P11} for all MHIMh models. Regardless of the scale, they lead to ratios that are within 10\% and those differences lowers with $\mu$. On linear scales, the effective bias gets closer to the linear one as $\mu$ increases. We can also notice in the lower panel that RSD effects impact the power spectrum only at $\mu>0.2$. On smaller scales, the rise is due to non-linear effects, only, at $\mu=0$ and also to RSDs for $\mu>0$. The slope of the rise scales with $\mu$ and it is exactly over the BAO scale range.
\subsection{A scale dependent HI bias on BAO scales}
\begin{figure}
\includegraphics[scale=0.97]{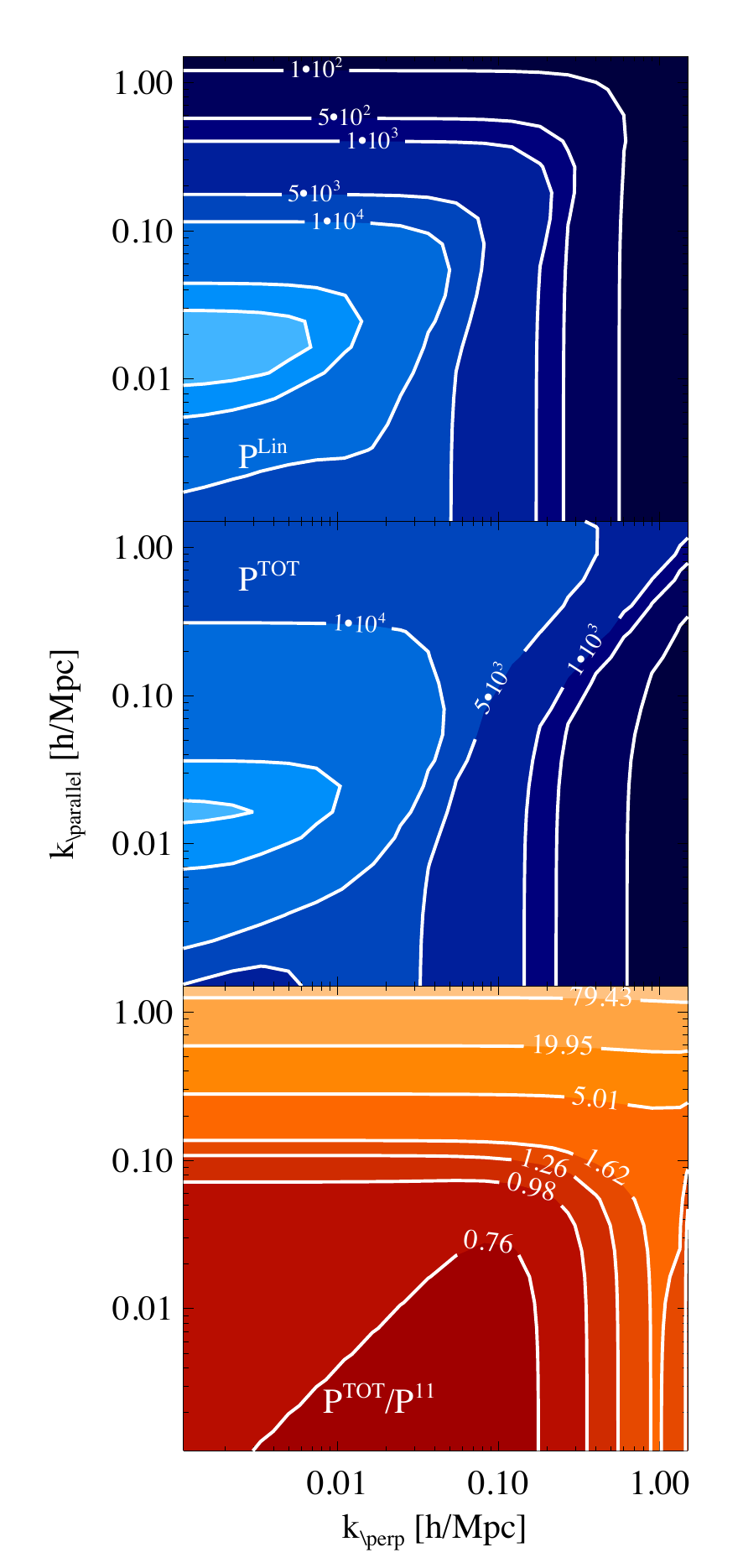} 
\caption{The anisotropic HI power spectrum for the HOD A model at $z=1$ broken down into transverse and radial directions (in $(\mbox{Mpc}\, h^{-1})^3$). The top and middle panels display the linear and total power spectra, respectively, while the bottom panel shows the ratio between both power spectra.}
\label{fig:PHI_kpp}
\end{figure}

We will carry on the analysis using only the model HOD A. The bottom panel of Fig. \ref{fig:ratio_PHI_P11} shows a scale dependence of the HI bias that is enhanced by RSD effects over the BAO scale range : the bias rises from 10 \% at $\mu=0$ to a factor 2 at $\mu=1$. To adopt an observational point of view, we change the coordinates to transverse and radial directions in Fig.~\ref{fig:PHI_kpp}. The two top panels show the linear and total HI power spectra.  At first glance, non-linear terms shift the turnover of the power spectrum towards higher $k_\perp$ ($>0.1\,h\, $Mpc$^{-1}$). They slightly enhance the signal along the $k_\perp$ direction while it is boosted along $k_\parallel$ by RSD effects. The lower panel of Fig. \ref{fig:PHI_kpp} shows the ratio between the linear and total HI power spectra. On large scales, $k_\parallel, k_\perp<0.01\,h\, $Mpc$^{-1}$, we recover a maximum ratio of 25\%. At small $k_\parallel$ and towards large $k_\perp$, non-linearities increase the amplitude of the HI power spectrum by a factor 2 while at large $k_\parallel$ RSD effects dominate non-linear ones and make any $k_\perp$ dependence vanish. Over the BAO scale range, the HI power spectrum increases by a factor of 5 and  2 in the radial and transverse directions, respectively. Therefore, both non-linearities and RSD effects modify the ratios between the BAO peaks. It is therefore necessary to take non-linearities into account when estimating cosmological parameters. To circumvent the contamination by non linear effects, one would preferentially measure the BAO peaks in the transverse direction but \citet{2017MNRAS.466.2736V} showed that, in single dish mode, beyond a certain size, the beam of the instrument smears the wiggles out in the transverse direction and that BAOs can only be detected in the radial direction. This is a limitation for the SKA and Meerkat but not for BINGO, CHIME or HIRAX as they will have a higher angular resolution. \\
%
%
\section{Conclusion}\label{par:ccl}
Radio telescopes are about to open a new window of observation on the Universe, in particular, using 21 cm intensity mapping. We investigate the non-linear power spectrum of HI in both real and redshift space and in the light of the relation between the halo mass and the HI mass.\\
Our main result is that on linear scales, the HI bias is \textit{not} constant but scale-dependent. Using a full 1-loop development in perturbation theory of the power spectrum of HI we show that non-linear contributions remove power to the HI power spectrum on linear scales in both real and redshift space at $z=1$. This result is contrary to our expectations and is not found in other modeling approaches. Commonly, a distribution of dark matter halos is `painted' with a baryonic tracer, so that only non-linearities coming from the distribution of dark matter are taken into account and not the ones coming from the evolution of the distribution of the tracer. \\
In real space, the effective bias of HI is 10-15\% lower than the linear one, depending on the MHIMh relation. The assumption that the observed HI bias is linear underpredicts the actual linear bias, and hence, the mass of halos hosting HI. \\
In redshift space, the effective bias is also lower than its linear counterpart up to 15\% and its scale dependence is highly sensitive to RSD effects. Over the BAO scale range, the HI bias rises with a slope that steepens with $\mu$. Regardless of the MHIMh prescription, the difference between the linear and the full HI power spectra reaches a factor 5 which can lead to a modification of the ratios between BAO peaks. Therefore, it will be crucial to take non-linearities into account when estimating cosmological parameters. \\
The different MHIMh relations lead to variations of 15\% at most on the HI bias. It is within the error bars on any of the current HI bias measurements so it is not an issue at the moment. Nevertheless, it will be indispensable for the upcoming HI surveys. It is worth noting that the observable is the product $T_\HI\, b_\HI$ where the HI temperature is also a function of the  MHIMh relation through $\Omega_\HI$. This product differs up to a factor of 7 between the different prescriptions.\\
Thorough forecasts of the effect of non-linearities on the estimation of BAO peaks and in a broader way, cosmological parameters, are required, including the redshift evolution as HI is positively biased at higher redshift, therefore non-linearities add power to the power spectrum of HI. Lastly, this effect is not only present in HI intensity mapping surveys but in any galaxy surveys. 
\section*{Acknowledgments}
The authors would like to thank Roy Marteens, Chris Clarkson, Jose Fonseca, and Vincent Desjacques for useful discussions. They also would like to thank Debanjan Sarkar for providing them the data points of their paper. AP, OU and MGS acknowledge support from the South African Square Kilometre Array Project as well as from the National Research Foundation.

%
%

\appendix
\section{Perturbation Theory formulas}\label{app:ir_int}
The necessary kernels are :  $F_2$ the non-linear density kernel, $G_2$ induced by peculiar velocities at second order, and $K_R$ arises from non-linear mode coupling  \citep[velocity-velocity and velocity-density,][]{2002PhR...367....1B}
\begin{IEEEeqnarray}{rCl}
F_2(\k_1,\k_2) &=& \frac{5}{7} + \frac{1}{2} \frac{\k_1\cdot\k_2}{k_1k_2} \left[ \frac{k_1}{k_2} + \frac{k_2}{k_1}\right] + \frac{2}{7} \left[  \frac{\k_1\cdot\k_2}{k_1k_2} \right]^2\\
G_2(\k_1,\k_2) &=& \frac{3}{7} + \frac{1}{2} \frac{\k_1\cdot\k_2}{k_1k_2} \left[ \frac{k_1}{k_2} + \frac{k_2}{k_1}\right] + \frac{4}{7} \left[  \frac{\k_1\cdot\k_2}{k_1k_2} \right]^2\\
K_R(\k_1,\k_2) &=& f\, b_1 \mu_1^2 + f\,b_1\, \mu_2^2 \nonumber \\
&+&\mu_1\mu_2 \left[  f\, b_1\frac{k_1}{k_2} +f\, b_1 \frac{k_2}{k_1}  \right]\nonumber\\
&+&f^2 \left[  2\mu_1^2\,\mu_2^2 + \mu_1\mu_2\left( \mu_1^2\frac{k_1}{k_2}  + \mu_2^2\frac{k_2}{k_1}  \right)    \right]
\end{IEEEeqnarray}
%
%
The matter and velocity power spectra at third order write 
\bea 
P_\m^{13}(k) &=& \frac{1}{252}\frac{k^3}{4\pi^2} P_\m^{11}(k) \int_0^\infty \dd r\,  P_\m^{11}(kr)\Big[ \frac{12}{r^2} - 158 \\
&+& 100\, r^2 - 42\, r^4 + \frac{3}{r^3}(r^2 - 1)^3(7r^2 + 2) 
\log\left|  \frac{1+r}{1-r} \right|\Big]\nonumber \\
P_\theta^{13}(k) &=&\frac{1}{84}\frac{k^3}{4\pi^2} P_\m^{11}(k) \int_0^\infty \dd r\,  P_\m^{11}(kr)\Big[ \frac{12}{r^2} - 82 \nonumber\\
&+& 4\, r^2 - 6\, r^4 + \frac{3}{r^3}(r^2 - 1)^3(r^2 + 2) 
\log\left|  \frac{1+r}{1-r} \right|\Big]
\eea
The last component of the $P_\HI^{13}(k,\mu)$ is
\begin{eqnarray}
I_R(k,\mu)&=& \frac{k^3}{(2\pi)^2} \int \dd r P^{11}(kr)\\ \nonumber 
& \times& \mu^2 f (\left[b_2B_1(r) + b_1 B_2(r) \right]          \\ \nonumber
&+&         \mu^2 f^2 \left[ b_1 B_3(r) + B_4 + \mu^2(b_1^2B_5(r) + fB_6(r))\right])
 \end{eqnarray}
with 
\begin{eqnarray}\nonumber
B_1(r) &=& \frac{1}{6} \\ \nonumber
B_2(r) &=&\frac{1}{84} \left[ -2 (9 r^4 - 24 r^2 + 19) + \frac{9}{r}(r^2 - 1) \log\left(\frac{1+r}{|1-r|}\right)\right]\\ \nonumber
B_3(r) &=& -\frac{1}{3}\\ \nonumber
B_4(r) &=&-\frac{1}{336\, r^3} [ 2(-9\, r^7 + 33\,r^5 + 33\, r^3 - 9\, r ) \\ \nonumber 
         &+& 9(r^2 - 1) \log\left(\frac{1+r}{|1-r|}\right)]\\ \nonumber
B_5(r) &=& \frac{1}{336\, r^3}[ 2\, r (-27\, r^6 + 63\, r^4 - 109\, r^2 + 9) \\ \nonumber
         &+& 9 (3\, r^2 +1)(r^2 - 1) \log\left(\frac{1+r}{|1-r|}\right)] \nonumber
 \end{eqnarray}
Lastly, in this framework, the full 1-loop matter power spectrum in real space is 
\be 
P_\mathrm{m}^\mathrm{NL}(k) = P_\mathrm{m}^{11}(k) + P_\mathrm{m}^{22}(k) +P_\mathrm{m}^{13}(k) 
\ee
where the second order term writes 
\be
P_\mathrm{m}^{22}(k) = \frac{1}{2}\int \frac{\dd^3 k_1}{(2\pi)^3} F_2^2(\k_1,\k_2) P_\m^{11}(k_2)\, P_\m^{11}(k_1)
\label{eq:Pm_real}
\ee
In redshift space, the full 1-loop matter power spectrum is 
\bea 
P_\mathrm{m}^\mathrm{NL}(k,\mu) &=& P_\mathrm{m}^{11}(k,\mu) + P_\mathrm{m}^{22}(k,\mu) +P_\mathrm{m}^{13}(k,\mu) \\
P_\mathrm{m}^{11}(k,\mu) &=& \left[1 + f\, \mu^2\right]^2\, P_\m^{11}(k)\\
P_\mathrm{m}^{22}(k,\mu) &=& \frac{1}{2}\int \frac{\dd^3 k_1}{(2\pi)^3} \Big[ F_2(\k_1,\k_2) \nonumber \\
 &+& \mu^2 G_2(\k_1,\k_2) + K_R(\k_1,\k_2)  \Big]^2 \nonumber \\
&\times& P_\m^{11}(k_2)\, P_\m^{11}(k_1)\\
P_\mathrm{m}^{13}(k,\mu) &=& \left(1 + \mu^2 f\right) \Big\{ I_R(k,\mu)\, P_\m^{11}(k) \nonumber\\
&+& \left[ P^{13}_\m(k) + \mu^2\,f\, P^{13}_\theta(k) \right]\Big\}
\label{eq:Pm}
\eea

\bibliographystyle{mn2e}
\bibliography{Biblio}

\label{lastpage}

\end{document}